# The study of Lorenz and Rössler strange attractors by means of quantum theory

**Yu I Bogdanov[1,2,3], N A Bogdanova[2]**

[1]Institute of Physics and Technology, Russian Academy of Science, 117218, Nakhimovskii pr. 36/1, Moscow, Russia
[2]National Research University of Electronic Technology MIET, 124498, Bld. 5, Pas. 4806, Zelenograd, Russia
[3]National Research Nuclear University "MEPHI", 115409, Kashirskoe highway 31, Moscow, Russia

E-mail: bogdanov_yurii@inbox.ru

**Abstract**

We have developed a method for complementing an arbitrary classical dynamical system to a quantum system using the Lorenz and Rössler systems as examples. The Schrödinger equation for the corresponding quantum statistical ensemble is described in terms of the Hamilton-Jacobi formalism. We consider both the original dynamical system in the position space and the conjugate dynamical system corresponding to the momentum space. Such simultaneous consideration of mutually complementary position and momentum frameworks provides a deeper understanding of the nature of chaotic behavior in dynamical systems. We have shown that the new formalism provides a significant simplification of the Lyapunov exponents calculations. From the point of view of quantum optics, the Lorenz and Rössler systems correspond to three modes of a quantized electromagnetic field in a medium with cubic nonlinearity. From the computational point of view, the new formalism provides a basis for the analysis of complex dynamical systems using quantum computers.

## 1. Representation of classical dynamical systems by means of quantum statistical ensembles

Classical dynamical system is described by means of an autonomous system of differential equations:

$$\frac{dx_j}{dt} = F_j\left(x_1,...,x_n\right) \ , \ j = 1,...,n \tag{1}$$

Formally, we can assume that the system of equations (1) defines the motion of a particle in an *n*-dimensional phase space. Values $F_j\left(x_1,...,x_n\right)$ specify components of the particle's velocity as a function of the coordinates.

As an illustration, we consider two important dynamical systems: the Lorenz system [1] and the Rössler system [2]. In both cases, the point moves in a three-dimensional space $\left(x_1, x_2, x_3\right)$ .

Velocity components for the Lorenz system are:

$$F_1 = -\sigma x_1 + \sigma x_2 \ , \ F_2 = r x_1 - x_2 - x_1 x_3 \ , \ F_3 = -b x_3 + x_1 x_2 \ , \tag{2}$$

where $\sigma = 10$, $b = 8/3$, $r = 28$.

Velocities components for the Rössler model are:

$$F_1 = -x_2 - x_3,\ F_2 = x_1 + a x_2,\ F_3 = b + x_1 x_3 - c x_3, \tag{3}$$

where $a = 0.2$, $b = 0.2$, $c = 5.7$.

Description in terms of statistical ensembles means that we consider the continuity equation for the density distribution in the phase space. This equation can be written as the Liouville's equation:

$$i\frac{\partial \rho}{\partial t} = L\rho, \qquad L = -iF_j \frac{\partial}{\partial x_j} - i\frac{\partial F_j}{\partial x_j}. \tag{4}$$

Here $L$ – is the Liouville operator. Analogues of (4) are widely used in studies on the statistical mechanics. We should note the contribution of Ilya Prigogine to this subject [3].

Liouville equation in phase space (4) can be regarded as a consequence of the Schrödinger equation for the wave function:

$$i\frac{\partial \psi}{\partial t} = H\psi, \qquad H = -iF_j \frac{\partial}{\partial x_j} - \frac{i}{2}\frac{\partial F_j}{\partial x_j} = \frac{1}{2}\left(F_j P_j + P_j F_j\right). \tag{5}$$

Here $P_j = -i\frac{\partial}{\partial x_j}$ are the momentum operators in the position representation. Density is given by the usual formula of quantum mechanics: $\rho = |\psi|^2$.

It is important to note that the Hamiltonian $H$ in the Schrödinger equation (5) is Hermitian, while the Liouville operator in the equation for the density (4) is not Hermitian, if the divergence of the velocity is different from zero ( $div\vec{F} = \frac{\partial F_j}{\partial x_j} \neq 0$ ).

Equation (5) that takes the form of the Schrödinger equation can be considered as the basis for a natural extension of the formalism of dynamical systems to the quantum domain. From a computational point of view, the solution of equation (5) can be built on entirely new hardware based on quantum computers. From the point of view of quantum optics, Lorenz and Rössler attractors correspond to strongly squeezed states of the electromagnetic field.

## 2. Hamilton-Jacobi formalism

The Schrödinger equation (5) can be considered the same as the Hamilton-Jacobi equation, if we represent it in the following form:

$$\frac{\partial \psi}{\partial t} + \mathcal{H} = 0, \qquad \mathcal{H} = F_j \frac{\partial \psi}{\partial x_j} + \frac{1}{2}\frac{\partial F_j}{\partial x_j}\psi. \tag{6}$$

Below, we assume that the psi-function is real. If the wave function is complex-valued, then the equation (6) splits into two independent equations.

Let us introduce canonical momenta

$$p_j = \frac{\partial \psi}{\partial x_j}. \tag{7}$$

Then the Hamiltonian function is of the form:

$$\mathcal{H}(x,p) = F_j(x)\,p_j + \frac{1}{2}\frac{\partial F_j(x)}{\partial x_j}\psi \,. \tag{8}$$

Canonical equations for the dynamical system will be:

$$\frac{dx_j}{dt} = \frac{\partial \mathcal{H}}{\partial p_j} = F_j(x_1,\dots,x_n), \tag{9}$$

$$\begin{aligned} \frac{dp_j}{dt} &= -\frac{\partial \mathcal{H}}{\partial x_j} = -\frac{\partial F_k}{\partial x_j}p_k - \frac{1}{2}\frac{\partial}{\partial x_j}\left(\left(\frac{\partial F_k}{\partial x_k}\right)\psi\right) = \\ &= -\frac{\partial F_k}{\partial x_j}p_k - \frac{1}{2}\left(\frac{\partial F_k}{\partial x_k}\right)p_j - \frac{1}{2}\frac{\partial}{\partial x_j}\left(\left(\frac{\partial F_k}{\partial x_k}\right)\right)\psi \end{aligned} \,. \tag{10}$$

The equations for the canonical coordinates (9) coincide with the original equation of the dynamic system (1). The equations for the canonical momenta (10) give us additional information that characterizes the evolution of the field gradient $\frac{\partial \psi}{\partial x_j}$.

In accordance with the Lagrangian description, consider the total (the substantial) derivative of the wave function $\psi$. This value characterizes the rate of change of the field in the coordinate system associated with the considered moving point. Taking into account (6) we obtain:

$$\frac{d\psi}{dt} = \frac{\partial \psi}{\partial t} + \frac{dx_j}{dt}\frac{\partial \psi}{\partial x_j} = -\frac{1}{2}\frac{\partial F_j}{\partial x_j}\psi \,. \tag{11}$$

Let us consider the logarithmic gradient of the field

$$\pi_j = \frac{1}{\psi}\frac{\partial \psi}{\partial x_j} = \frac{p_j}{\psi} \,. \tag{12}$$

The entered value allows us to represent the evolution of the gradient in the "pure" form, getting rid of the explicit dependence of the field on time. Using (10) and (11), we obtain the following equation for the dynamics of the logarithmic gradient of the field

$$\frac{d\pi_j}{dt} = B_{jk}\pi_k - \frac{1}{2}\frac{\partial}{\partial x_j}\left(\frac{\partial F_k}{\partial x_k}\right), \tag{13}$$

where

$$B_{jk} = -\frac{\partial F_k}{\partial x_j} \,. \tag{14}$$

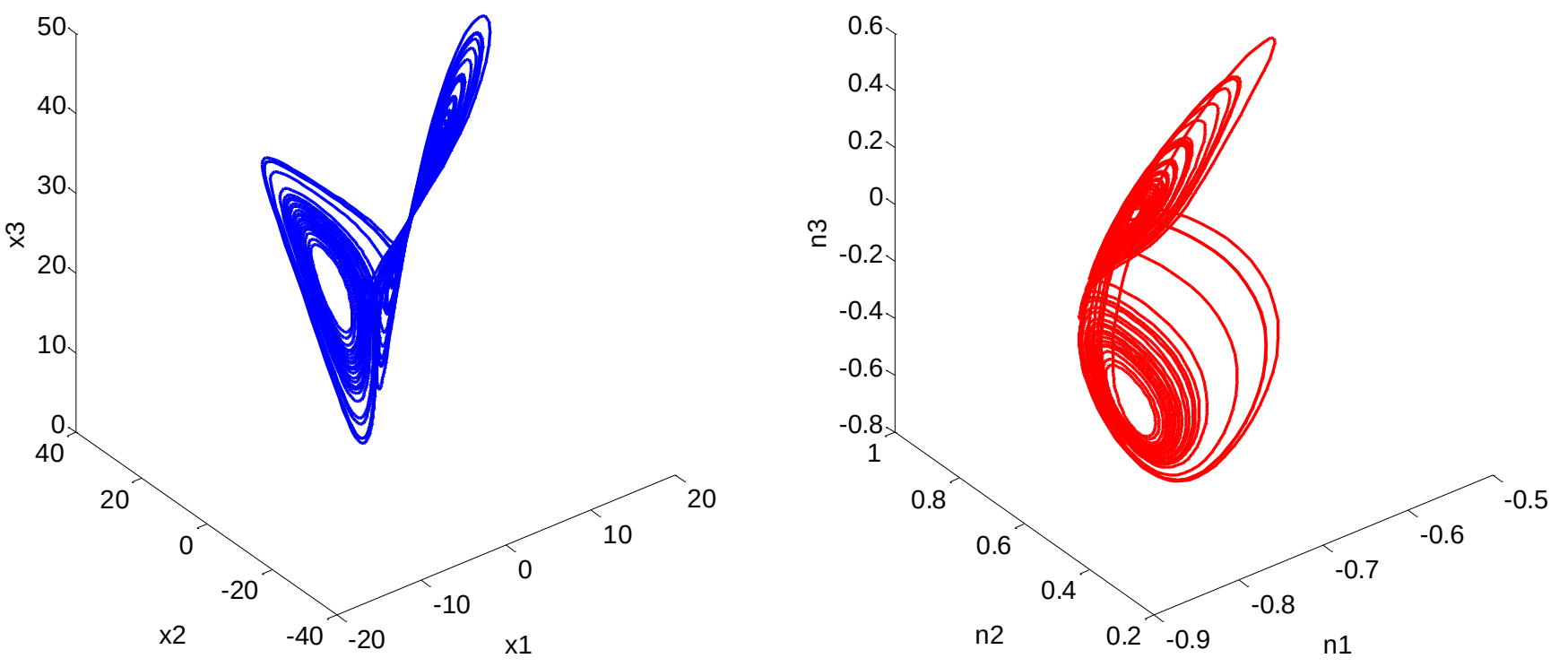

Figure 1. Attractor in the Lorenz system. Mutually complementary position and momentum pictures

The figure 1 shows a set of mutually complementary pictures for the Lorenz system: positions (left) and momentum directions (right). Directions of the momentum indicate the directions of maximum compression (squeezing) in the phase space.

The following figure 2 shows the direction of the momentum in spherical coordinates for a particle moving on the Lorenz attractor.

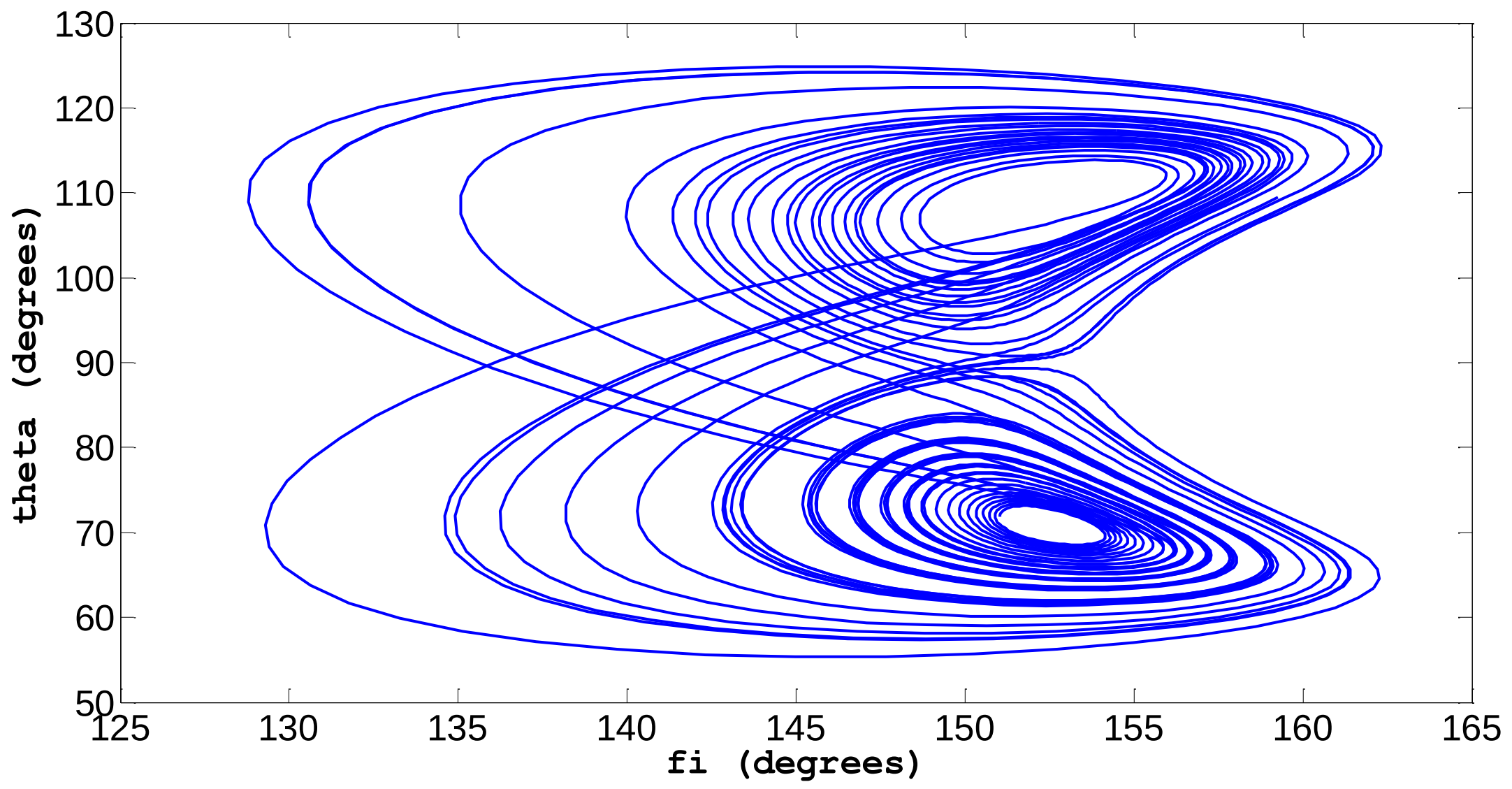

Figure 2. Attractor in the Lorenz system. Momentum picture. Plane (fi, theta)

The figure 3 is similar to figure 1 and shows a set of mutually complementary pictures for the Rössler system: positions (left) and momentum directions (right).

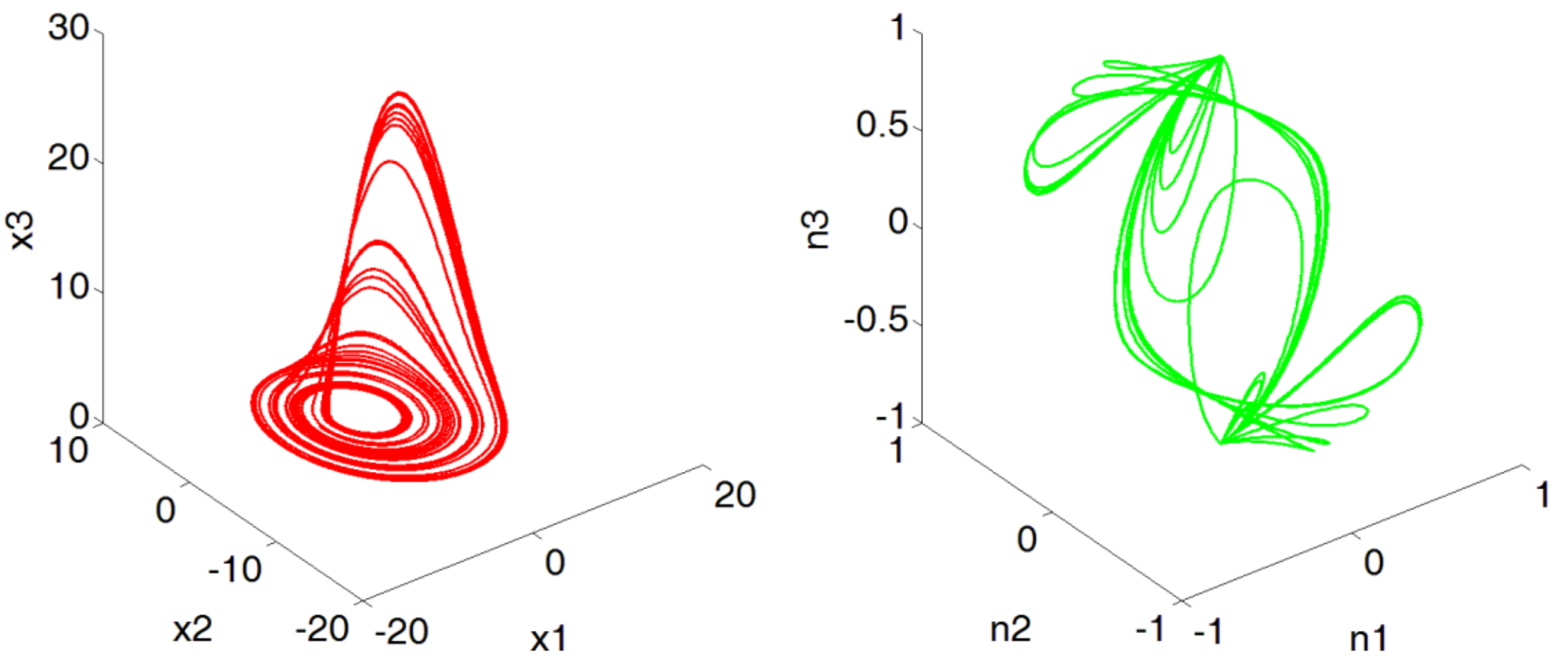

Figure 3. Attractor in the Rössler system. Mutually complementary position and momentum pictures

The figure 4 is similar to figure 2 and shows the direction of the momentum in plane (n1, n2) for the Rössler system .

The following figure 5 illustrates the exponentially rapid contraction of the phase space in the Lorenz system. It is assumed that at time zero there is a wide spherically symmetric Gaussian distribution with standard deviation equal to sig = 40. As a result of compression, the distribution quickly takes the shape of the Lorenz attractor in the form of "butterfly wings." figure 5 considers the density distribution in the direction of maximum compression. The distribution within a "butterfly wing" at time points t = 0.6 (left) and t = 1.0 (right) is shown. The initial coordinates of the point are x = (10,10,10). We see that during dt = 0.4 the “wing” is compressed about 300 times.

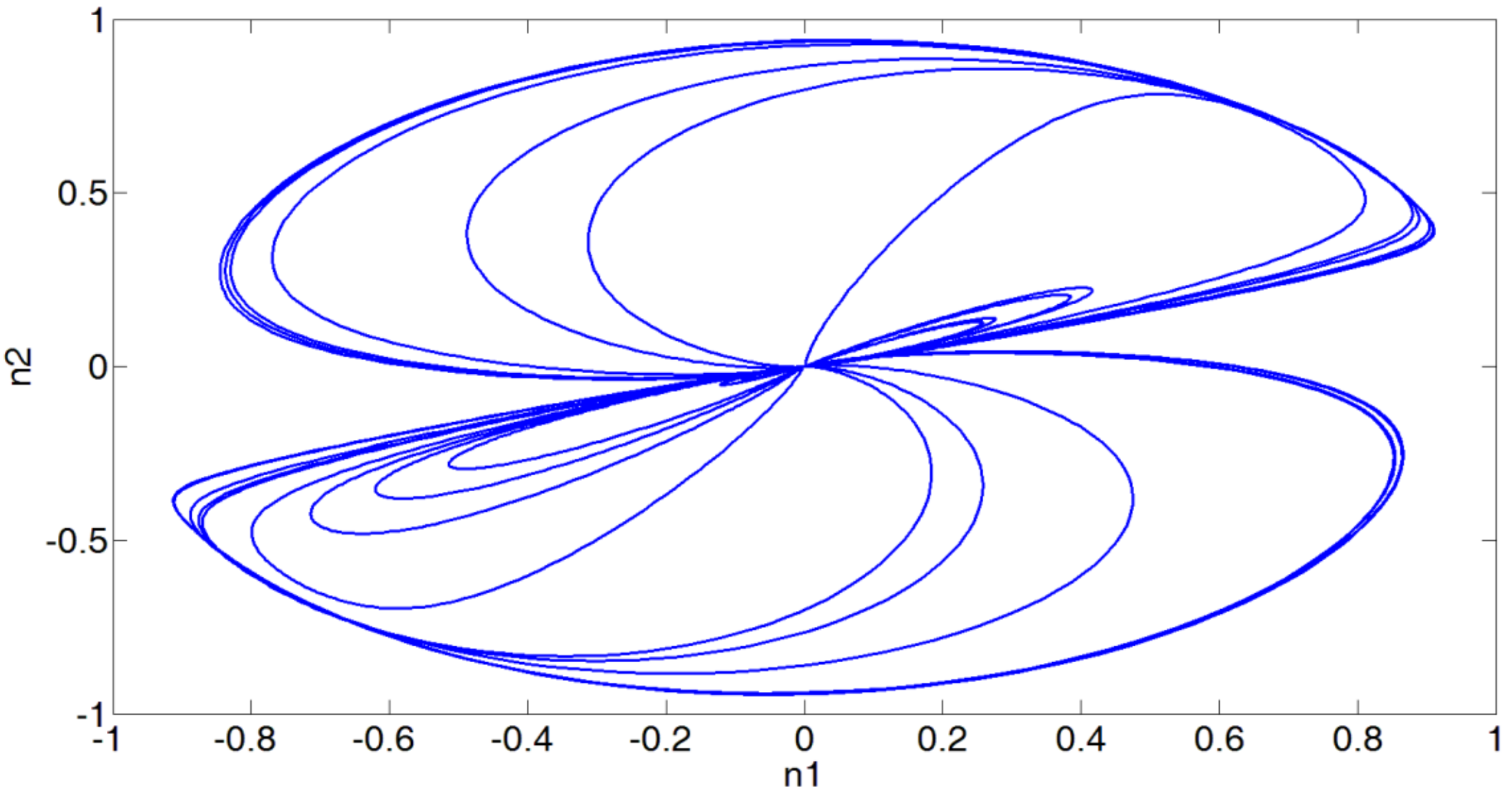

Figure 4. Attractor in the Rössler system. Momentum picture. Plane (n1,n2)

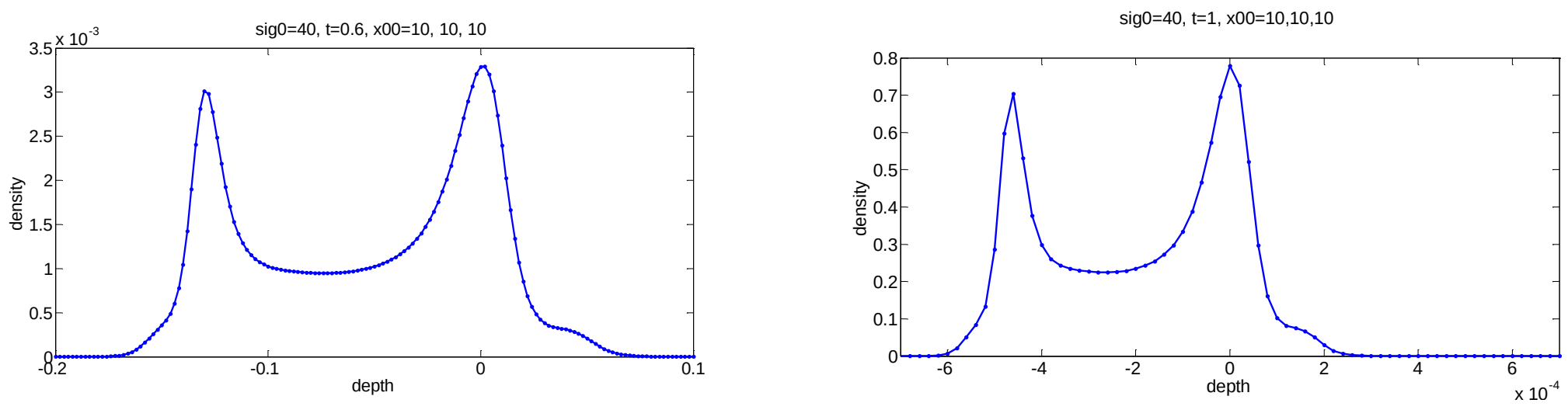

Figure 5. The evolution of the thickness of the “butterfly's wing”. Density distribution inside the “wings of a butterfly” at time t = 0.6 (left) and at time t = 1.0 (right)

## 3. Calculation of Lyapunov exponents

Consider the application of the above approach to the strange attractors appearing in the Lorenz and Rössler systems. Because of the dissipativity of considered systems, the sum of the Lyapunov exponents must be negative. Due to the condition, the dissipative attractor is an attracting set of measure zero in the phase space. All points, regardless of their initial position, approach the attractor exponentially fast and concentrate near it over time.

Let us consider three orthogonal directions: the direction of maximum compression of the phase space $\vec{n} = \dfrac{\vec{\pi}}{|\vec{\pi}|}$, direction of the velocity $\vec{v} = \dfrac{\vec{V}}{|\vec{V}|}$, where $\vec{V} = \left(F_1, F_2, F_3\right)$ and finally, the direction given by the unit vector $\vec{\tau}$, which is orthogonal to the two former directions.

Note that in the direction $\vec{v}$ the phase volume is not compressed nor expanded on average, so that the corresponding Lyapunov exponent is zero.

In the direction $\vec{\tau}$ the phase volume expands on average that corresponds to a positive Lyapunov exponent.

Finally, in the direction $\vec{n}$ the phase volume contracts (shrinks) on average that corresponds to a negative Lyapunov exponent.

We introduce the projections corresponding to the three directions given above:

$$P_n = |n\rangle\langle n| = \begin{pmatrix} n_1 \\ n_2 \\ n_3 \end{pmatrix} \begin{pmatrix} n_1 & n_2 & n_3 \end{pmatrix},$$

$$P_v = |v\rangle\langle v| = \begin{pmatrix} v_1 \\ v_2 \\ v_3 \end{pmatrix} \begin{pmatrix} v_1 & v_2 & v_3 \end{pmatrix},$$

$$P_\tau = |\tau\rangle\langle \tau| = \begin{pmatrix} \tau_1 \\ \tau_2 \\ \tau_3 \end{pmatrix} \begin{pmatrix} \tau_1 & \tau_2 & \tau_3 \end{pmatrix},$$

These projections form an orthogonal decomposition of unity:

$$P_n + P_v + P_\tau = I .$$

Here $I$ is the identity matrix of size $3 \times 3$.

Lyapunov exponents in the momentum representation are defined by the average value of the operator $B$ (14).

The negative sign in the formulas below corresponds to the transition to the position representation

$$\lambda_3 = \lambda_n = -\langle\langle n|B|n\rangle\rangle = -\langle Tr(BP_n)\rangle, \tag{15}$$

$$\lambda_2 = \lambda_v = -\langle\langle v|B|v\rangle\rangle = -\langle Tr(BP_v)\rangle, \tag{16}$$

$$\lambda_1 = \lambda_\tau = -\langle\langle \tau|B|\tau\rangle\rangle = -\langle Tr(BP_\tau)\rangle. \tag{17}$$

Here, the symbol "$Tr$", as customary, stands for the trace of the matrix. Angular brackets denote the averaging procedure in time as the point moves to the attractor. Averaging in time is equivalent to averaging over the ensemble (ergodicity).

The results of numerical calculations of the Lyapunov exponents for the Lorenz attractor are as follows:

$$\lambda_1 = 0.90410 \pm 0.00076,\ \lambda_2 = -0.000082 \pm 0.00017,\ \lambda_3 = -14.57068 \pm 0.00075.$$

Our values are close to the values obtained by Sprott J.C. [4] and Kuznetsov S.P. [5]:

$$\lambda_1 = 0.906,\ \lambda_3 = -14.572, \qquad \text{(Sprott J. C.)}$$

$$\lambda_1 = 0.897,\ \lambda_3 = -14.563. \qquad \text{(Kuznetsov S.P.)}$$

The results of numerical calculations of the Lyapunov exponents for the Rössler attractor are the following:

$$\lambda_1 = 0.07062 \pm 0.00066, \ \lambda_2 = 0.000048 \pm 0.00018, \ \lambda_3 = -5.3937 \pm 0.0018.$$

Our values are close to the values obtained by Sprott J.C. [4]:

$$\lambda_1 = 0.0714, \ \lambda_3 = -5.3943, \qquad \text{(Sprott J. C.)}$$

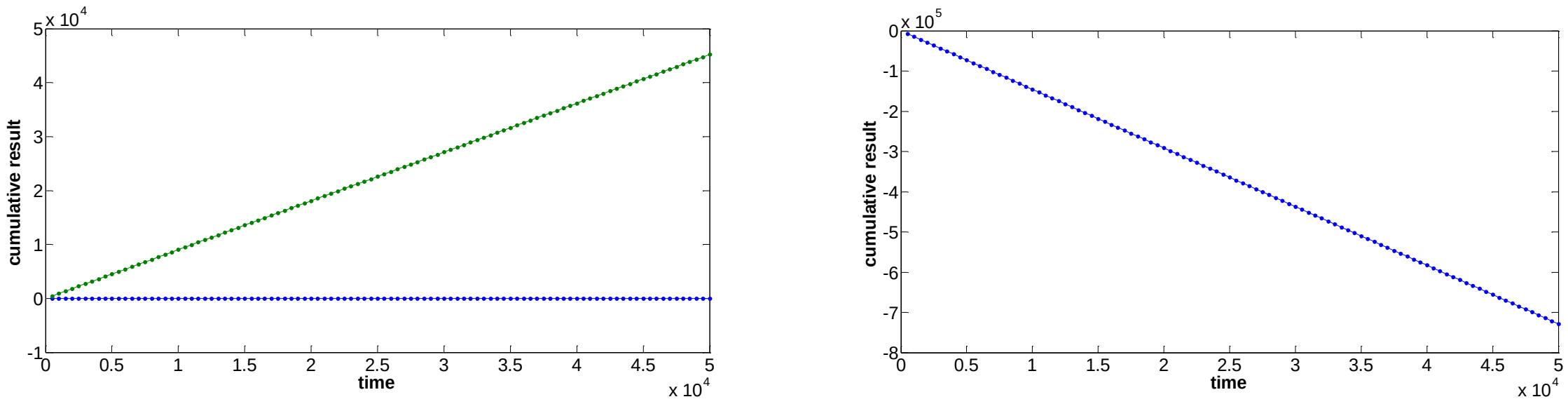

Figure 6. The accumulated data for the Lorenz attractor. The first and the second Lyapunov exponents (left), the third Lyapunov exponent (right).

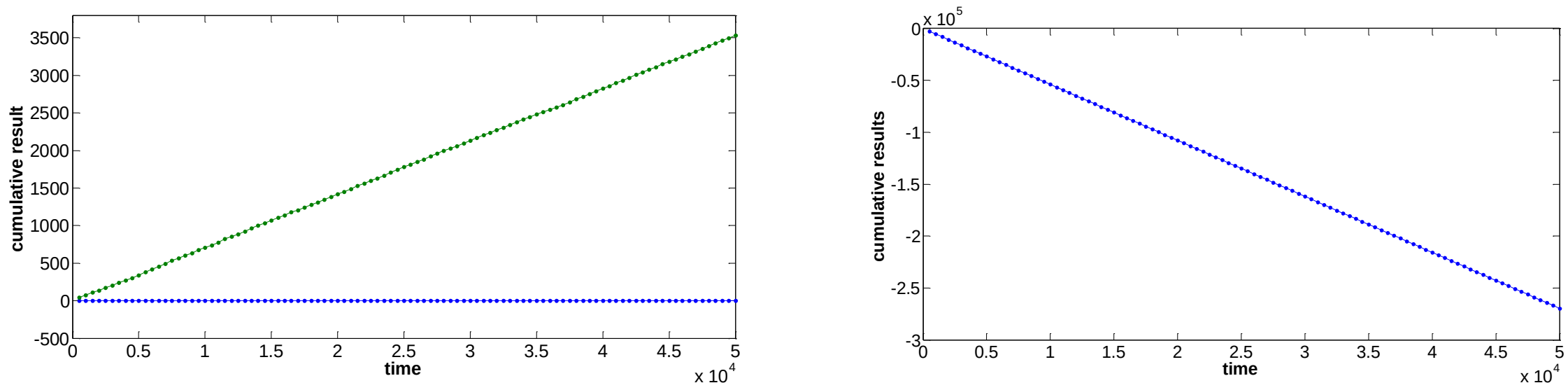

Figure 7. The accumulated data for the Rössler attractor. The first and the second Lyapunov exponents (left), the third Lyapunov exponent (right).

The procedure for the statistical estimation of Lyapunov exponents is graphically presented in figure 6 and figure 7. Here, the Lyapunov exponents are determined by the slope of the straight lines corresponding to the accumulated statistics.

Lorenz and Rössler strange attractors are fractals. The Kaplan- Yorke fractal dimension is given by the following formula

$$d_{KY} = 2 + \frac{\lambda_1}{|\lambda_3|}.$$

In our case, we obtain the following approximate values

$$d_{KY} = 2.062, \qquad \text{(Lorenz strange attractor)}$$

$$d_{KY} = 2.013. \qquad \text{(Rössler strange attractor)}$$

## 4. Conclusions

We have developed a method for complementing an arbitrary classical dynamical system to a quantum system using the Lorenz and Rössler systems as examples.

The Schrödinger equation for the corresponding quantum statistical ensemble is described in terms of the Hamilton-Jacobi formalism.

Simultaneous consideration of mutually complementary position and momentum frameworks provides a deeper understanding of the nature of chaotic behavior in dynamical systems.

We have shown that the new formalism provides a significant simplification of the Lyapunov exponent calculations.

From the point of view of quantum optics, the Lorenz and Rössler systems correspond to three modes of a quantized electromagnetic field in a medium with cubic nonlinearity.

From the computational point of view, the new formalism provides a basis for the analysis of complex dynamical systems using quantum computers. Generation of three-photon and multi-photon states in a nonlinear medium is a core technology to create full-scale quantum computing devices.

## Acknowledgements

The work was supported by the Russian Science Foundation, grant № 14-12-01338.